\documentclass[12pt]{iopart}

\usepackage{color}

\begin{document}

\title{The span of correlations in dolphin whistle sequences}

\author{Ramon Ferrer-i-Cancho$^{1,*}$ and Brenda McCowan$^{2}$} 

\address{$^1$ Complexity \& Quantitative Linguistics Lab \\
Departament de Llenguatges i Sistemes Inform\`atics, \\
TALP Research Center, Universitat Polit\`ecnica de Catalunya, \\
Campus Nord, Edifici Omega Jordi Girona Salgado 1-3. \\
08034 Barcelona, Catalonia (Spain)}

\address{$^2$ Population Health \& Reproduction, \\ 
School of Veterinary Medicine, \\
University of California, \\ 
One Shields Ave 1024 Haring Hall \\
Davis, CA 95616, United States
}

\eads{\mailto{rferrericancho@lsi.upc.edu} and \mailto{bjmccowan@ucdavis.edu}}

\begin{abstract}
Long-range correlations are found in symbolic sequences from human language, music and DNA. Determining the span of correlations in dolphin whistle sequences is crucial for shedding light on their communicative complexity. Dolphin whistles share various statistical properties with human words, i.e. Zipf's law for word frequencies (namely that the probability of the $i$th most frequent word of a text is about $i^{-\alpha}$) and a parallel of the tendency of  more frequent words to have more meanings. The finding of Zipf's law for word frequencies in dolphin whistles has been the topic of an intense debate on its implications. One of the major arguments against the relevance of Zipf's law in dolphin whistles is that is not possible to distinguish the outcome of a die rolling experiment from that of a linguistic or communicative source producing Zipf's law for word frequencies. Here we show that statistically significant whistle-whistle correlations extend back to the 2nd previous whistle in the sequence using a global randomization test and to the 4th previous whistle using a local randomization test. None of these correlations are expected by a die rolling experiment and other simple explanation of  Zipf's law for word frequencies such as Simon's model that produce sequences of unpredictable elements.
\end{abstract}




\maketitle

\section{Introduction}

Long-range correlations have been reported in different kinds of symbolic sequences: human language \cite{Moscoso2011a, Montemurro2011a, Montemurro2001b, Ebeling1995a}, DNA \cite{Li1992} and music \cite{Su2007a, Ebeling1995a}. 
A few studies have studied the span of correlations in sequences of behavior produced by other species \cite{Ferrer2005h, McCowan1999, Altmann1965a}.
Rather long-correlations (extending back at least to the 7th previous element) have reported in sequences of dolphin surface behavioral patterns \cite{Ferrer2005h}. A preliminary analysis of constraints in sequences of dolphin whistles was performed in Ref. \cite{McCowan1999} but strong conclusions were not reached due to the small size of the dataset. Determining the actual span of correlations in dolphin sequences is crucial for shedding light on the communicative complexity of dolphins whistles. 

Various similarities between human words and dolphin whistles have been reported. A parallel of Zipf's law of meaning distribution, the tendency or more frequent words to have more meanings \cite{Zipf1949a}, has been found in dolphins whistles \cite{Ferrer2009f}. 
Zipf's law for word frequencies, namely, the probability of the $i$th most frequent word of a text, is $\sim i^{-\alpha}$, where $\alpha$ is the exponent of the law \cite{Zipf1949a}, has also been found in dolphin whistles \cite{McCowan1999}. However, the finding remains controversial because it has been argued that simply rolling a die
could explain the presence of the law in dolphin whistles \cite{Suzuki2004a}. The experiment consists of generating a sequence of faces by rolling a die. One of the faces of the die plays the role of a word delimiter. For instance, a die of six faces could produce 1, 5, 2, 6, 3, 2, 4, 3, 5, 6... Treating $6$ as a pseudo-word delimiter, the previous sequence of faces becomes the sequence of pseudo-words $152, 32435,...$. The experiment is an abstraction of the popular monkey typing experiment, that consists of typing at random on a keyboard (the face that plays the role of the word delimiter is the space and the other faces are letters) \cite{Miller1963}. Die rolling and monkey typing have been argued to explain or mimic Zipf's law for word frequencies in human words \cite{Miller1963} and dolphin whistles \cite{Suzuki2004a}.
Another die rolling experiment, where the probability of the $i$th face is the probability of the $i$th most frequent word, has been proposed \cite{Niyogi1995a}.

The hypothesis of die rolling as an explanation for Zipf's law for word frequencies in human language in other species can be tested in at least two different ways: 
\begin{enumerate}
\item
{\em By comparing the actual distribution of 'word' frequencies with the one that is actually produced by the die rolling process \cite{Ferrer2009b}.} Concerning the die rolling experiment of Ref. \cite{Suzuki2004a}, the parameters of the model that provide a satisfactory fit to actual word frequencies according to a statistically rigorous test, are indeed unknown \cite{Ferrer2009b}, in spite of the many previous claims about the good fit of the model using qualitative arguments \cite{Miller1963, Suzuki2004a}. Concerning the die rolling experiment or Ref. \cite{Niyogi1995a}, the model is able to trivially provide a perfect fit to any theoretical or empirical discrete distribution, being Zipf's law for word frequencies a particular case.  
\item
{\em By comparing the statistical properties of the sequence of words produced with those of the sequence that is produced by the die rolling process.} The rolling experiments of Refs. \cite{Suzuki2004a, Niyogi1995a} produce a sequence of independent 'words' in the sense that elements that have already been produced carry no information about the next element. 
In contrast human language, shows long range correlations in texts using words (e.g. \cite{Montemurro2011a, Montemurro2001b}) or letters (e.g., \cite{Moscoso2011a, Ebeling1994}) as units of the sequence.
\end{enumerate}
Here we will follow the second track for dolphin whistles. The aim of the present article is to determine the span of correlations in dolphin whistle sequences and evaluate the suitability of die rolling \cite{Suzuki2004a, Niyogi1995a}. The challenge of the analysis is facing the statistical problems arising from the rather small size of the dataset of Ref. \cite{McCowan1999}. The danger of undersampling in the context of dolphins whistles has already been discusssed \cite{McCowan1999}. The next section presents the information theoretic measure approach that will be used to study correlations in dolphin whistle sequences in that dataset.

\section{Mutual information}

We consider pairs of whistles in a sequence and three related random variables: $X$ for the whistle that is the 1st member of the pair, $Y$ for the whistle that is the second member of the pair and $D$ for their distance. We adopt the convention that consecutive elements are at distance $1$, elements separated by one element are at distance $2$, and so on...\cite{Ferrer2004b}.
Given a collection of sequences of whistles, $p(X = x , Y = y | D = d)$ is defined as the probability that whistle $x$ is followed by whistle $y$ knowing that they are at distance $d$ from each other and $d_{max}$ is defined as the maximum distance considered in the analysis (thus $1 \leq d \leq d_{max}$).
Given a certain distance $d$, the marginal conditional probabilities are defined as
\begin{eqnarray*}
p(X = x | D = d) & = & \sum_y p(X = x, Y = y | D = d) \\
p(Y = y | D = d) & = & \sum_x p(X = x, Y = y | D = d).
\end{eqnarray*}
$I(X; Y | D = d)$, the conditional mutual information between $X$ and $Y$ given a concrete distance $d$, is defined as \cite{Cover2006a}
\begin{equation}
\fl I(X; Y | D = d) = \sum_{x, y} p(X = x, Y = y | D = d) \log q(X = x, Y = y | D = d),
\label{mutual_conditional_entropy_given_concrete_distance_equation}
\end{equation}
where \begin{equation*}
q(X = x, Y = y | D = d) = \frac{p(X=x, Y=y | D = d)}{p(X = x | D = d)p(Y = y | D = d)}.
\end{equation*}
$I(X; Y | D = d)$ has been used to study long-range correlations in DNA, texts and music \cite{Li1992,Ebeling1995a}.

$f(X = x, Y = y | D = d)$ is defined as the number of times that $x$ has been followed by $y$ at distance $d$. The marginal conditional
frequencies are defined as
\begin{eqnarray}
f(X = x | D = d) & = & \sum_{y} f(X = x, Y = y | D = d) \label{marginal_left_frequency} \\
f(Y = y | D = d) & = & \sum_{x} f(X = x, Y = y | D = d), \label{marginal_right_frequency}
\end{eqnarray}
and the total number of pairs at distance $d$ is defined as  
\begin{equation*}
F(d) = \sum_{x} f(X = x | D = d) = \sum_{y} f(Y = y | D = d).
\end{equation*}
In a finite collection of sequences, $p(X=x, Y=y | D = d)$, $p(X = x | D = d)$ and $p(Y = y | D = d)$ are relative frequencies, i.e. 
\begin{eqnarray}
p(X = x, Y = y | D = d) & = & \frac{f(X = x, Y = y | D = d)}{F(d)} \label{relative_joint_frequency_equation} \\
p(X = x | D = d) & = & \frac{f(X = x | D = d)}{F(d)} \label{relative_left_marginal_frequency_equation} \\
p(Y = y | D = d) & = & \frac{f(Y = y | D = d)}{F(d)} \label{relative_right_marginal_frequency_equation}.
\end{eqnarray}
In a real collection of sequences, $p(X=x, Y=y | D = d)$, $p(X = x | D = d)$ and $p(Y = y | D = d)$ are estimated from these relative frequencies.
Applying Eqs. \ref{relative_joint_frequency_equation}, \ref{relative_left_marginal_frequency_equation} and \ref{relative_right_marginal_frequency_equation} to Eq. \ref{mutual_conditional_entropy_given_concrete_distance_equation}, yields the {\em sample} mutual information between $X$ and $Y$ given a concrete distance $d$,
\begin{equation}
I_s(X; Y | D = d) = \log F(d) + \frac{\sigma_1 - \sigma_2 - \sigma_3}{F(d)},
\label{sample_mutual_information_equation}
\end{equation}
where 
\begin{eqnarray*}
\sigma_1 & = & \sum_{x,y} f(X = x, Y = y | D = d) \log f(X = x, Y = y | D = d) \\
\sigma_2 & = & \sum_x f(X = x | D = d) \log f(X = x | D = d) \\
\sigma_3 & = & \sum_y f(Y = y | D = d) \log f(Y = y | D = d).
\end{eqnarray*}
Next two useful properties of $I_s(X; Y | D = d)$ are presented:
\begin{enumerate}
\item
If marginal conditional frequencies at distance $d$ are boolean, i.e. $f(X = x | D = d) = f(Y = y | D = d) \in \{ 0, 1 \}$ for any $x$ and $y$ then $I_s(X; Y | D = d) = \log F(d)$. To see it, notice that $f(X = x | D = d) = f(Y = y | D = d) = 1$ implies, thanks to Eqs. \ref{marginal_left_frequency} and \ref{marginal_right_frequency}, that the joint frequencies are all boolean, i.e. $f(X = x, Y = y | D = d) \in \{0, 1 \}$ for any $x$ and $y$. According to Eq. \ref{sample_mutual_information_equation}, the fact that both marginal and joint conditional frequencies do not exceed one yields $\sigma_1 = \sigma_2 = \sigma_3 = 0$ and thus $I_s(X; Y | D = d) = log F(d)$ as we wanted to prove.
\item
If all whistles at cooccurring at distance $d$ are identical (only one whistle type has non-zero frequency), then $\sigma_1 = \sigma_2 = \sigma_3 = F(d) \log F(d)$ and then, according to  Eq. \ref{sample_mutual_information_equation}, $I_s(X; Y | D = d) = 0$.
\end{enumerate}

\section{Methods}

We reused the collection of whistle sequences employed to study Zipf's law in dolphin whistles \cite{McCowan1999}. A summary of the elementary statistical properties of the sequences is provided in Table \ref{summary_table}. Sequences of length smaller than $2$ where filtered out. 

\subsection{Distances where correlations are significant}

For each dolphin in the dataset, his/her collection of sequences was analyzed to extract a list of distances in the interval $[2, d_{max}^2]$ at which $I_s(X; Y | D = d)$ is statistically significant at a significance level of $a = 0.05$. For each dolphin and each distance, we used a Monte Carlo procedure for estimating a $p$-value indicating the probability that the value of $I_s(X; Y | D = d)$ from a randomized version of the data $I_s(X; Y | D = d)$ is at least as large as that of the original collection of sequences:
\begin{enumerate}
\item
$R = 10^7$ randomized versions of the original data were generated. 
\item 
$R_{\geq}$, the number of times the value of $I_s(X; Y | D = d)$ is at least as large as that of the original data was calculated for each $d \in [1, d_{max}^2]$ and the $p$-value was estimated as $R_{\geq}/R$. 
\item 
For each dolphin, all distances $d$ such that $p$-value $= R_{\geq}/R \leq a$ were added to the list of distances.
\end{enumerate}

\subsection{Upper bounds for the span of correlations}

We say that $I_s(X ; Y | D = d)$ is constant if $I_s(X ; Y | D = d)$ is the same for any randomization of data upon which $I_s(X ; Y | D = d)$ is computed. We consider two situations in which $I_s(X; Y | D = d)$ cannot be significantly high: (a) $I_s(X ; Y | D = d)$ is constant in the sense above or (b) $I_s(X; Y; D = d) = 0$ as $0$ is actually the minimum of mutual information \cite{Cover2006a}. Accordingly, we consider three different ways of defining $d_{max}$:
\begin{itemize}
\item
$d_{max}^0$, defined through the number of pairs of elements at distance $d$ in a sequence of length $l$, i.e. $\pi(d, l) = l - d$. The maximum distance $d$ that can be considered in a collection of sequences where the longest sequence has length $l_{max}$ is obtained from $\pi(d_{max}, l_{max}) \geq 1$, which gives $d_{max}^0 = l_{max} - 1$.
\item
$d_{max}^1$, defined as the smallest distance at which $I_s(X ; Y | D = d)$ is constant in the sense above for any $d \in (d_{max}^1, d_{max}^0]$. 
\item
$d_{max}^2$, defined as the smallest distance beyond which $I_s(X; Y | D = d)$ cannot be significantly high for $d \in (d_{max}^2, d_{max}^0]$. Notice that $d_{max}^2 \leq d_{max}^1$ because $I_s(X ; Y | D = d)$ cannot be significantly high if $I_s(X ; Y | D = d)$ is constant.
\end{itemize}
The interest of these definitions of $d_{max}$ is two-fold. First, bounding a priori the span of correlations between whistles. Second, reducing the computational cost of evaluating the significance of $I_s(X ; Y | D = d)$ at distances where the result of the test is straightforward. Distances greater than $d_{max}^2$ can be discarded.
 
We did  not check if $I_s(X; Y | D = d) = 0$ using Eq. \ref{sample_mutual_information_equation} as this is problematic due to finite numerical precision for real numbers. Eq. \ref{mutual_conditional_entropy_given_concrete_distance_equation} indicates that $I_s(X; Y | D = d) = 0$ if and only if $q(X = x; Y = y | D = d) = 1$ for any $x$ and $y$ such that $p(X = x; Y = y | D = d) > 0$. Applying Eqs. \ref{relative_joint_frequency_equation}, \ref{relative_left_marginal_frequency_equation} and \ref{relative_right_marginal_frequency_equation}, it is easy to see that the condition $q(X = x; Y = y | D = d) = 1$ is equivalent to a more numerically convenient condition, i.e. $F(d) f(X = x, Y = y | D = d) = f(X = x | D = d) f(Y = y | D = d)$. 

\subsection{Kinds of randomization}

Two kind of randomizations of the data upon $I_s(X; Y | D = d)$ is calculated were considered: global and local randomization. A precise calculation of $d_{max}^1$ and $d_{max}^2$ depends on the kind of randomization.

\begin{table}
\caption{\label{summary_table} Summary of the elementary statistical properties of the collections of sequences of each dolphin. For each dolphin, the following information is shown: dolphin's name, $T$, the total number of whistle types, $V$, the number of different whistle types, $S$ (the number of sequences), $\left<l \right>=T/S$ (the mean sequence length in whistle types), $l_{max}$ (the maximum length). Dolphins are sorted decreasingly by $T$. } 
\begin{indented}
\item[]\begin{tabular}{@{}llllll}
\br
Name & $T$ & $V$ & $S$ & $\left< l \right>$ & $l_{max}$ \\
\hline
\mr
Liberty & 334 & 31 & 102 & 3.27 & 11 \\
Norman & 235 & 39 & 74 & 3.18 & 21 \\
Panama & 148 & 42 & 43 & 3.44 & 10 \\
Chelsea & 110 & 12 & 34 & 3.24 & 8 \\
Sadie & 110 & 13 & 33 & 3.33 & 8 \\
Stormy & 94 & 12 & 27 & 3.48 & 9 \\
Delphi & 65 & 18 & 24 & 2.71 & 7 \\
Neptune & 53 & 9 & 9 & 5.89 & 13 \\
Circe & 47 & 4 & 15 & 3.13 & 7 \\
Desmond & 40 & 3 & 7 & 5.71 & 10 \\
Sam & 39 & 4 & 12 & 3.25 & 13 \\
Tasha & 35 & 9 & 12 & 2.92 & 7 \\
Bayou & 27 & 10 & 8 & 3.38 & 8 \\
Terry & 21 & 7 & 5 & 4.20 & 7 \\
Schooner & 19 & 4 & 6 & 3.17 & 6 \\
ECB & 15 & 5 & 6 & 2.50 & 4 \\
Gordo & 6 & 6 & 2 & 3.00 & 4 \\
\mr
\end{tabular}
\end{indented}
\end{table}

\subsubsection{Global randomization}

Here the data that is randomized is the whole collection of sequences. 
The randomization procedure consisted of copying all whistles in a vector of length
\begin{equation*}
T = \sum_{i = 1}^S l_i, 
\end{equation*}
where $l_i$ is the length of the $i$th sequence and $S$ is the number of sequences of the collection. Then, every randomized version of the collection is obtained by generating a uniformly distributed random permutation of the vector \cite{Durtenfeld1964a} and cutting that vector in pieces of lengths $l_1,...,l_i,...l_S$ (cut always in this order) to produce the randomized collection of sequences. Notice that the randomization procedure preserves the sequence lengths and the frequencies of each whistle in the original collection of sequences. Concerning the computation of $d_{max}^1$, notice that constant $I_s(X ; Y | D = d)$ with regard to any global randomization can occur in three circumstances: 
\begin{enumerate}
\item 
If all the whistle types are hapax legomena, i.e. they occur only once in the whole collection as all the marginal conditional frequencies are boolean and thus $I(X ; Y | D = d) = \log F(d)$, as it has been shown above, for any $d \in [1, d_{max}^0]$. 
\item  
If the repertoire size is one (there is only one whistle type of non-zero frequency) then all the whistles cooccurring at a certain distance are identical and thus $I(X ; Y | D = d) = 0$, as it has been shown above, for any $d \in [1, d_{max}^0]$.
\item
When $F(d) = 1$, all the marginal conditional frequencies are boolean in this case and then $I(X ; Y | D = d) = \log F(d) = \log 1 = 0$. 
\end{enumerate}
These conditions lead to the following specific procedure for calculating $d_{max}^1$.
Satisfying condition (i) (this is the case of the dolphin 'Gordo') or (ii) implies $d_{max}^1 = 0$. If none of these two conditions is met, $d_{max}^1$ is calculated by means of condition (iii). In order to warrant $F(d) > 1$, $d_{max}^1$ must be $d_{max}^0$ if there is more than one sequence of maximum length and $d_{max}^1 = d_{max}^0 - 1$ otherwise. To see it, notice that if the collection has $S_{max}$ sequences of maximum length, then $F(d_{max}^1) = S_{max} \pi(l_{max} -1, l) = S_{max}$ and thus $F(d) \geq 2$ if and only if $S_{max} \geq 2$.

\subsubsection{Local randomization}

Here the data that is randomized are the pairs of whistles occurring at a certain distance $d$.
The randomization procedure consisted of copying all the pairs of whistles occurring at distance $d$ in a vector of length $2F(d)$.
Then, every randomized version of the collection was obtained by generating a uniformly distributed random permutation of the vector \cite{Durtenfeld1964a} and then taking the $i$th and the $(i+1)$th whistles of the vector, with $1 \leq i \leq 2F(d)$, to form the $j$th pair of whistles of the randomized pairs of whistles, with $1 \leq j \leq F(d)$ and $j= \lceil i/2 \rceil$. Notice that the randomization procedure preserves the frequencies of each whistle in the original pairs of whistles at distance $d$.

Concerning the computation of $d_{max}^1$ for local normalization notice that constant $I_s(X ; Y | D = d)$ with regard to any local randomization for distance $d$, adds two new relevant conditions with regard to global randomization: 
\begin{enumerate}
\item[(iv)]
When all whistle types are locally hapax legomena, i.e. all the whistle types forming the pairs at distance $d$ occur only once in the ensemble of pairs at distance $d$. In that case, all marginal conditional frequencies are boolean and thus $I_s(X; Y; D = d) = F(d)$ as it has been shown above.
\item[(v)]
When the local repertoire has size one, i.e. all the pairs at distance $d$ are made of a single whistle type. In that case, all  $I_s(X; Y; D = d) = 0$ as it has been shown above.
\end{enumerate}
The procedure for calculating $d_{max}^1$ under local normalization is the following. First, compute $d_{max}^1$ with the same procedure used for global normalization. Decrease $d_{max}^1$ while $d_{max}^1 >0$ and at least one of the two following conditions is met at distance $d_{max}^1$: (a) all whistle types are locally hapax legomena or (b) the local repertoire size is one. 

\begin{table}
\caption{\label{span_table}Analysis of the span of correlation between whistles at a certain distance $d$ using global and local randomization.
For the collection of whistles sequences of each dolphin, the following information is shown: the dolphin's name, 
$d_{max}^{0,1,2}$ (the maximum distance according to different definitions) and the list of distances $d \in [1, d_{max}^2]$ such that $I_s(X; Y | D = d)$ is significantly high at a significance level of 0.05. Dolphins are sorted decreasingly by $T$. 
} 
\begin{indented}
\item[]\begin{tabular}{@{}lllllllllllll}
\br
   & & \centre{3}{Global randomization} & \centre{3}{Local randomization} \\ 
\ns
   & & \crule{3} & \crule{3} \\
Name & $d_{max}^0$ & $d_{max}^1$ & $d_{max}^2$ & Distances & $d_{max}^1$ & $d_{max}^2$ & Distances \\ 
\hline
\mr
Liberty &  10 & 9 & 7 &  &  8 & 7 & 1, 4, 6, 7 \\
Norman &  20 & 19 & 15 & 1, 2 &  19 & 15 & 1, 2, 3, 4 \\
Panama &  9 & 8 & 7 & 5 &  8 & 7 & 1, 2, 3 \\
Chelsea &  7 & 6 & 3 & 1 &  4 & 3 & 1, 2 \\
Sadie &  7 & 6 & 6 & 1, 2 &  6 & 6 & 1, 2 \\
Stormy &  8 & 7 & 6 & 1 &  7 & 6 & 1, 2, 3 \\
Delphi &  6 & 5 & 5 &  &  4 & 4 & 1 \\
Neptune &  12 & 11 & 7 & 2 &  11 & 7 & 1, 2, 3, 4 \\
Circe &  6 & 5 & 1 &  &  1 & 1 &  \\
Desmond &  9 & 8 & 3 & 1 &  8 & 3 &  \\
Sam &  12 & 11 & 9 & 7 &  10 & 9 &  \\
Tasha &  6 & 5 & 5 &  &  5 & 5 &  \\
Bayou &  7 & 6 & 6 &  &  5 & 5 & 1, 2 \\
Terry &  6 & 5 & 3 & 1 &  4 & 3 & 1 \\
Schooner &  5 & 4 & 2 & 1 &  4 & 2 & 1 \\
ECB &  3 & 2 & 2 &  &  2 & 2 &  \\
Gordo &  3 & 0 & 0 &  &  0 & 0 &  \\
\mr
\end{tabular}
\end{indented}
\end{table}

\section{Results}

\label{results_section}

Table \ref{span_table} shows the values of $d$ where $I_s(X; Y | D = d)$ is significantly high in the dolphins from Ref. \cite{McCowan1999} at a significance level of 0.05 for global and local randomization. 
$n_d^a$ is defined as the number of dolphins for which $I_s(X; Y | D = d)$ is significantly high at a significance level $a$. Table \ref{span_table} yields $n_1^{0.05} = 7$, $n_2^{0.05} = 3$, $n_5^{0.05} = n_7^{0.05} = 1$ for global randomization and $n_1^{0.05} = 11$, $n_2^{0.05} = 7$, $n_3^{0.05} = 4$, $n_4^{0.05} = 3$, $n_6^{0.05} = n_7^{0.05} = 1$ for local randomization. 

Next we will perform a meta-analysis to determine when $n_{d}^{0.05}$ is significantly high because $n_{d}^{0.05}$ could be the outcome of false positives (type I statistical errors) from the test
that determines if $I_s(X; Y | D = d)$ is significantly high for a certain dolphin. The $p$-value of a continuous statistic is known to be uniformly distributed under the null hypothesis \cite{Rice2007a}. In our case, $I_s(X; Y | D = d)$ is approximately continuous and the quality of the approximation is expected to increase with the size of the permutation  space of the collection of sequences. This implies that $n_d^a$, the number of dolphins for which $I_s(X; Y | D = d)$ is significantly high at a significance level $a$ follows approximately a binomial distribution with parameters $N_d$ and $a$, where $N_d$ is the total number of dolphins for whom $I_s(X; Y | D = d)$ can be significantly high according to the criteria above, and $a = 0.05$ is the significance level. Notice that the dolphin 'Gordo' is a special case because he does no have distances at which mutual information can be significantly high (he has $d_{max}^1 = d_{max}^2 = 0$ because all the whistles in his collection are hapax legomena; notice $T = V$ in Table \ref{summary_table}). Thus, $N_d$ is not simply the number of dolphins in the dataset. 

A binomial test can be used to asses if $n_d^a$ is significantly large. The $p$-value of this test is
\begin{equation}
p-value = \sum_{x = n_d^a}^{N_d} {N_d \choose x} a^{x} (1 - a)^{N_d - x}.
\label{p_value_equation}
\end{equation}
Concerning global randomization, a binomial test indicates that the number of dolphins showing a significant correlation is significantly high for $d = 1$ and $d = 2$ but not for $d > 2$ (Table \ref{meta_analysis_table}).
Concerning local randomization, a binomial test indicates that the number of dolphins showing a significant correlation is significantly high for $1\leq d \leq 4$ but not for $d > 4$ (Table \ref{meta_analysis_table}). 

\begin{table}
\caption{\label{meta_analysis_table} Summary of the test of the significance of $n_d^{0.05}$, the number of dolphins for which $I_s(X; Y | D = d)$ is significantly high at a significance level of $0.05$. Distances where $n_d^{0.05}$ cannot be significantly high (as $n_d^{0.05} = 0$ is minimum) are excluded from the analyses and filled with $-$. $N_d$ is the number of dolphins for whom $I_s(X; Y | D = d)$ can be significantly high. The $p$-values indicate the probability of reaching a value of $n_d^{0.05}$ at least as high as the actual one simply by chance. The $p$-values were estimated by means of two different procedures: a binomial approximation and a Monte Carlo method. $p$-values were rounded to leave only one significant digit. } 
\begin{indented}
\item[]\begin{tabular}{@{}lllllllll}
\br
   & \centre{4}{Global randomization} & \centre{4}{Local randomization} \\
\ns
   & \crule{4} & \crule{4} \\
   & $n_d^{0.05}$ & $N_d$ & \centre{2}{$p$-value}             & $n_d^{0.05}$ & $N_d$ & \centre{2}{$p$-value} \\
\ns
    &              &       & \crule{2}                      &              &       & \crule{2} \\
$d$ &              &       & Binomial & Numerical           &              &       & Binomial & Numerical \\
\hline
\mr
$1$ & $7$          & $16$  & $6 \times 10^{-6}$ & $<0.001$ & $11$         & $16$ & $2 \times 10^{-11}$ & $<0.001$ \\
$2$ & $3$          & $15$  & $0.04$             & $0.02$   & $7$          & $15$ & $3 \times 10^{-6}$  & $<0.001$ \\
$3$ & $0$          & $-$   & $-$                & $-$      & $4$          & $13$ & $0.003$             & $0.001$ \\
$4$ & $0$          & $-$   & $-$                & $-$      & $3$          & $10$ & $0.01$              & $0.009$ \\
$5$ & $1$          & $10$  & $0.4$              & $0.3$    & $0$          & $-$  & $-$                 & $-$ \\
$6$ & $0$          & $-$   & $-$                & $-$      & $1$          & $7$  & $0.3$               & $0.2$ \\ 
$7$ & $1$          & $5$   & $0.2$              & $0.2$    & $1$          & $5$  & $0.2$               & $0.2$ \\
\mr
\end{tabular}
\end{indented}
\end{table}

The meta-analysis based upon a binomial test assumes that 0.05 is the probability of rejecting the null hypothesis by chance but indeed the accuracy of the assumption depends on the properties of the collection of whistle types. For instance, we have seen that the mutual information cannot be significantly high for the dolphin 'Gordo' and thus the probability of rejecting the null hypothesis by chance is thus $0$ for him. A Monte Carlo meta-analysis allows one to improve the approximate calculation of the $p$-value offered by Eq. \ref{p_value_equation} to some degree of numerical precision. Consider that a randomized ensemble of collections of sequences consists of a randomized version of the collection of each of the dolphins (excluding 'Gordo'). For each distance $d$, the $p$-value of $n_d^a$ is estimated by the proportion of times in $R_m$ randomized ensembles of collections that the value of $n_d^a$ is equal or greater than the value of $n_d^a$ of the original (non-randomized) ensemble. $R_m = 1000$ and $R = 1000$ were used. 
Concerning global randomization, one obtains that $n_d^{0.05}$ is significantly high for $d = 1$ and $d = 2$ but not for $d > 2$  (Table \ref{meta_analysis_table}). 
Concerning local randomization, one obtains that $n_d^{0.05}$ is significantly high for $1 \leq d \leq 4$ but not for $d> 4$ (Table \ref{meta_analysis_table}). 
To sum up, the conclusions of the two meta-analyses coincide from a qualitative point of view (Table \ref{conclusions_meta_analysis_table}). The binomial test provides a good enough approximation for both global and local randomization. 

From a global randomization perspective, correlations are short range: the presence of significant correlations at low distances ($d=2$ and specially $d=1$) is unquestionable but significant correlations at higher distances could be false positives (Table \ref{conclusions_meta_analysis_table}). From a local randomization perspective, correlations extending back to the 4th back whistle type are unquestionable but farther significant correlations could be false positives (Table \ref{conclusions_meta_analysis_table}).  

\begin{table}
\caption{\label{conclusions_meta_analysis_table} Summary of the support for significance of each distance (based upon Table \ref{meta_analysis_table}).
$n_d^{0.05}$ is defined as the number of dolphins for which $I_s(X; Y | D = d)$ is significantly high at a significance level of $0.05$. For each kind of randomization (global or local) and meta-analysis (binomial approximation or Monte Carlo method), the distances at which $n_d^{0.05}$ is significantly high are indicated. }
\begin{indented}
\item[]\begin{tabular}{@{}lll}
\br
   & \centre{2}{Distances} \\ 
\ns
   & \crule{2} \\
   & Global randomization & Local randomization \\ 
\hline
\mr
$n_d^{0.05}>0$                         & $1,2,5,7$ & $1,2,3,4,6,7$ \\
Meta-analysis (binomial approx.) & $1,2$     & $1,2,3,4$ \\
Meta-analysis (Monte Carlo method)      & $1,2$     & $1,2,3,4$ \\
\mr
\end{tabular}
\end{indented}
\end{table}

\section{Discussion}

We have demonstrated that, for the majority of individuals, a dolphin whistle carries (on average) a significant amount of information about the next whistles of the sequence (Tables \ref{span_table} and \ref{meta_analysis_table}). Global randomization indicates that a whistle carries information about at least one of the next two whistles of the sequence whereas local randomization indicates that a whistle carries information about at least one of the next four whistles of the sequence (Table \ref{conclusions_meta_analysis_table}). This is a property that is inconsistent with die-rolling, where a pseudo-word carries no information at all about the next pseudo-words of the sequence. The fact that this is also a feature also shared by Simon's model for word frequencies \cite{Simon1955}, questions the validity of a popular argument against the utility of Zipf's law for frequencies, namely that the law is of practically no help in assessing the complexity of a communication system because there are many ways of reproducing it \cite{Miller1957, Rapoport1982, Niyogi1995a, Suzuki2004a}. 
Indeed, when the statistical properties of the sequence are taken into account, the number of candidate explanations drops down. The multiplicity of explanations for Zipf's law in a species such as dolphins depends on how many statistical features, besides Zipf's law for word frequencies, are used to break the tie between candidates. The big question that future research on dolphins whistles must address is: what is the communicative complexity of a system whose units (e.g., whistles types) are distributed following Zipf's law for word frequencies \cite{McCowan1999}, show a parallel of Zipf's law of meaning distribution \cite{Ferrer2009f} and form sequences with correlations that defy a simple explanation such as die rolling or Simon's model? We hope that our research stimulates further data collection to determine if the rather short range correlation discovered here are an intrinsic property of dolphin whistle communication or a consequence of the small size of our dataset. 

\ack 

We are grateful to L. Doyle for helpful discussions and to M. Piattelli-Palmarini for making us aware of Ref. \cite{Niyogi1995a}.
This work was supported by the grant {\em Iniciaci\'o i reincorporaci\'o a la recerca} from the Universitat Polit\`ecnica de Catalunya and the grants BASMATI (TIN2011-27479-C04-03) and OpenMT-2 (TIN2009-14675-C03) from the Spanish Ministry of Science and Innovation.

\section*{References}

\bibliographystyle{unsrt}

\end{document}